# Regulating the Energy Flow in a Cyanobacterial Light Harvesting Antenna Complex

Ido Eisenberg[1]*, Felipe Caycedo-Soler[2], Dvir Harris[3], Shira Yochelis[1], Susana F. Huelga[2], Martin B. Plenio[2], Noam Adir[3], Nir Keren[4], Yossi Paltiel[1]*

[1]Applied Physics Department and The Center for Nano-Science and Nano-Technology, The Hebrew University of Jerusalem, Jerusalem, 9190401 Israel
[2]Institute of Theoretical Physics, Ulm University, Albert Einstein Alle 11, 89069 Ulm, Germany
[3]Schulich Faculty of Chemistry, Technion - Israel Institute of Technology, Haifa, 32000 Israel
[4]Department of Plant and Environmental Sciences, Alexander Silberman Institute of Life Sciences, Givat Ram, The Hebrew University of Jerusalem, Jerusalem 9190401, Israel

Photosynthetic organisms harvest light energy, utilizing the absorption and energy transfer properties of protein-bound chromophores. Controlling the harvesting efficiency is critical for the optimal function of the photosynthetic apparatus. Here, we show that cyanobacterial light-harvesting antenna may be able to regulate the flow of energy in order to switch reversibly from efficient energy conversion to photo-protective quenching via a structural change. We isolated cyanobacterial light harvesting proteins, phycocyanin and allophycocyanin, and measured their optical properties in solution and in an aggregated-desiccated state. The results indicate that energy band structures are changed, generating a switch between two modes of operation: exciton transfer and quenching; achieved without dedicated carotenoid quenchers. This flexibility can contribute greatly to the large dynamic range of cyanobacterial light harvesting systems.

## INTRODUCTION

Photosynthetic organisms are able to thrive in environments in which the light intensities are constantly changing. This is made possible by the existence of molecular mechanisms that provide the dynamic range required for both extremely efficient excitation energy transfer (EET) to drive light-chemical energy conversion, and to quench energy when photochemistry is blocked or becomes rate-limiting.[1–3] In the green photosynthetic lineage (green algae and vascular plants) the EET process utilizes chlorophylls and energy quenching is performed mainly by carotenoids.[1,4–6] Balancing these two processes under changing external conditions is achieved by adaptation mechanisms that include structural and conformational changes.[1,7–10]

Within their antenna complex, the phycobilisome (PBS), cyanobacteria utilize linear tetrapyrrole (bilin) chromophores covalently bound to dedicated phycobiliproteins (PBPs) for EET.[7,11–13] In some species, under high light conditions, the orange carotenoid protein (OCP) is activated to promote quenching.[14–16] PBPs themselves do not contain any carotenoids.

The PBS is composed of PBPs. It has a central core which is made of 2-5 allophycocyanin (APC) cylinders, surrounded by 6-8 rods[7,17] (Figure 1a). All rods contain phycocyanin (PC) and occasionally, other hypsochromic variants of PBPs as well. This mega-structure architecture contains additional non-pigment binding proteins called linker proteins (Lp).[18] These Lp have a critical role in determining the directionality of EET[7].

PBPs are assembled from two types of subunits that form the (αβ) monomer. The PC monomer has three phycocyanobilin (PCB) chromophores: β155, α84 and β84[19–21] (Figure 1b) while APC contains only two PCBs: α84 and β84[22–25] (Figure 1c). Monomers further assemble into trimers, hexamers and finally into rod or core cylinders. These further assemble to form the complete PBS. Although PC and APC chromophores are chemically identical, their excited state energy levels are different, due to the influence of the surrounding protein.[26]

Photons are absorbed by PBP chromophores, creating an exciton. This exciton was suggested to hop from one chromophore to another by Förster Resonance Energy Transfer (FRET) or by stronger excitonic quantum coupling.[27–31] It migrates within and along the PC rods to the APC central core, finally reaching its terminal emitter and then to the reaction centers (RC).[6]

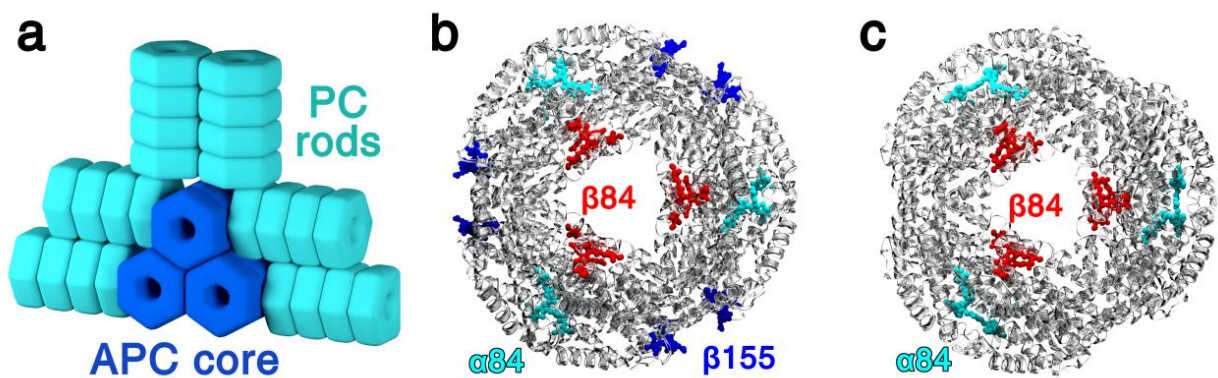

**Figure 1.** Cyanobacterial photosynthetic antenna proteins structure. (a) Schematic representation of typical tri-cylindrical phycobilisome light harvesting antenna, composed of an APC containing a core (dark blue) and PC rods surrounding the core (cyan). (b) Crystallographic derived phycocyanin hexamer structure contains 18 chromophores of three kinds: β155 (blue), α84 (cyan) and β84 (red). PDB ID: 3O18.[21] (c) Allophycocyanin hexamer contains 12 chromophores of two kinds: α84 (cyan) and β84 (red). PDB ID: 3DBJ.[24]

The PBS super-structure is highly adaptive to changing environmental conditions.[7,32–34] Although extensive efforts have been made to determine its ultra-structure by X-ray crystallography or electron-microscopy, so far only partial successes have been reported[7] and it is now clear that structural dynamics are of functional importance.

One example of a structural dynamic has been identified in desert sand crust cyanobacteria. This species can tolerate almost complete desiccation and sustain both hot and arid conditions by switching from a wet phase to a desiccated phase. In the latter phase, conformational changes to its PBSs were shown to occur and to be related to photosynthesis quenching mechanisms.[35]

In this research, we explore the optical properties of PC and APC isolated from the thermophilic cyanobacterium *Thermosynechococcus vulcanus* in two phases: wet and desiccated, to elucidate their EET mechanisms. We present both experimental evidence and theoretical calculations for two modes of operation: One that allows energy funneling towards the APC core, and another that induces energy quenching, utilizing the same photosynthetic chromophore.

## EXPERIMENTAL METHODS

The experimental procedure consists of isolation of PC and APC proteins, followed by re-suspension in six different buffer solutions (Table S1) and measurement of their optical properties in their completely wet phase and in the desiccated phase.

### Protein preparation

*T. vulcanus* cells were grown in a 2-liter temperature-controlled growth chamber in BG11 medium supplemented with 5% $CO_2$ in air at 55 C, with fluorescent lamp illumination. Cells were grown for 5 to 7 days before collection by centrifugation (6,000 × g for 15 min). The cells were re-suspended in isolation buffer (20 mM Tris, 10 mM $MgCl_2$ and 10 mM $CaCl_2$, pH 7.5) and treated with lysozyme (1 mg/ml) for 2 h at 50 C in the dark, before passing through a microfluidizer (M-110S) at maximal pressure. The sample was then centrifuged for 15 min at 1,200 × g to discard the unbroken cell fraction. The green supernatant was then incubated with 3% (vol/vol) Triton-X for 60 min at room temperature and centrifuged for 45 min at 27,500 × g. The supernatant containing the soluble PBP was separated using low pressure anion exchange chromatography (Toyopearl QAE-550C filled as resin). Fractions were characterized by absorption spectroscopy in order to select samples containing pure PBP. APC was obtained in high purity after this step. PC was dialyzed and further purified by high-pressure anion exchange column (Agilent PL-SAX 1000 Å 8 μ column). The resulting pure PC and APC were exchanged into 10 mM Potassium Phosphate (K-Pho), pH=7.5 using ultrafiltration (Millipore). The total volume of each sample was divided equally to yield 6 x 400 μl of PC at 10 mg/ml and 6 x 400 μl of APC at 3 mg/ml. Additional buffer exchanges were done in similar fashion for each sample, generating 6 samples of PC and 6 samples of APC containing different buffers (Table S1). One sample was dialyzed against DDW to achieve a very low salt concentration (25 μM). This sample is attributed as low salts (LS). Two other samples were dialyzed against 10 mM Tris HCL and Tris HBr buffers. The last three samples were dialyzed against 10 mM, 500 mM and 900 mM potassium phosphate buffer. In this article, all figures show data for LS samples except for Figure 3 and Figures S1-S4.

## Optical Measurements

Absorption measurements were performed using an integrating sphere. A white light source (ThorLabs QTH10) was coupled into a 150 mm diameter integrating sphere via a lens. The samples were then placed over the exit aperture. In the experimental setup the aperture was bigger than the sample. In this way the sample was equally illuminated from all directions by the integrating sphere. The transmitted light was collected by a lens coupled to an optic fiber that was connected to a visible range fiber spectrometer (OceanOptics USB4000). Spectrometer measurement error is <1%. Transmission spectra were normalized relative to clean glass substrate.

Luminescence measurements were performed with a different setup. A sample was illuminated by a 532 nm wavelength laser (Altechna DPSS CW GREEN Laser). In the optical setup the beam was expanded by lens and speckles were reduced by OptoTune Transmissive Laser Speckle Reducer. The excitation light was then filtered out by a 620 nm long-pass filter. Finally, the luminescence was collected by a lens coupled to an optic fiber that was connected into the same spectrometer as was used for the absorption measurements (OceanOptics USB4000).

The time-resolved fluorescent measurements were carried out using a time-correlated single-photon counting (TCSPC) approach. Samples were excited using a Fianium WhiteLase SC-400 super continuum laser monochromatized at 532 nm. Pulse FWHM was ~80 ps. The emission from the sample was collected by a 20X objective lens, through a long-pass filter and collected using a MPD PD-100-CTE-FC photon counter. We used the PicoHarp 300 as our TCSPC system.

All samples were prepared by dropping 4 μl of protein solution over a glass substrate. We measured both PC and APC, each of them in 6 different buffers (12 samples). Measurements were taken at room temperature. We repeated these measurements two times, for absorption and for luminescence, hence, overall we had 24 measurements. The drying process of the solution took 20-40 min, and we conducted measurements during the drying process. Every single measurement was integrated over 10 s, therefore we had 120-240 measurements. For life-time measurements we averaged every 6 consecutive measurements.

For absorbance and luminescence measurements, spectra changed sharply when the sample was dried. The dry phase was stable and we used the last spectrum as the dry spectrum. However, the wet phase showed some minor changes. We attribute these changes to the transition from trimer to hexamer. Late spectra in this phase were slightly narrower, hence we took the spectrum that had the smallest FWHM as the wet phase spectrum. This spectrum represents hexamer rather than trimer, with two exceptions being the two last buffers, 500 mM K-Pho and 900 mM K-Pho. These two buffers had high amounts of salts that retained high amounts of water in the samples. In these samples the changes from trimer to hexamer were not as prominent and we used the first spectrum as representative of the wet phase. All shown spectra are normalized to 1.

Life-time data was fitted using DecayFit 1.4 software. Every measurement was fitted with 2 exponents. Sample images were taken by a Motic PSM1000 optic microscope using CCD.

**RESULTS**

Measurements of absorption spectra of PC and APC show that in the desiccated ordered aggregate the absorption spectrum is broader compared to the wet phase, especially for PC (Figures 2a and S1-S2). In addition, the typical additional peak of the APC absorption at 650 nm is lost and the protein absorbs only at 620 nm. This has been previously shown to occur when the protein environment of the APC chromophores is altered.[24]

Measurements of the concomitant fluorescence spectra show a shift towards lower energies as a result of drying (Figures 2b and S3-S4). In their wet phase, the PC emission peak has higher energy than APC, enabling efficient energy transfer in the PBS. However, in their dry phase, the order is reversed and the APC peak is higher than that of PC. While in the wet phase two energetic levels contribute to the luminescence spectra, in the dry phase only one level is dominant. We also observed a significant reduction in the fluorescence intensity of about 98% and 96% for PC and APC, respectively, in the dry phase (Figures S5b, S5d and S5e).

Fluorescence life-time measurements were best fitted by two exponential decaying contributions (Figure 2c and Table S2). Both PC and APC exhibit shortened emission life-times in the dry phase, while the relative weights of the two contributions also varied from wet phase to dry phase (Figure S6). The fluorescence reduction is also demonstrated by the smaller area under the dry life-time traces. It is important to note that the process was reversible. Upon re-wetting the sample, reconstitution of the original spectral properties of the isolated proteins was achieved (Figures S5a and S5c).

Absorption and luminescence measurements of all six buffers (see table S1) were taken to study the influence of the buffer on the spectral results (Figure 3). In both PC and APC the buffer had relatively small influence in the wet phase and greater influence in the dry phase.

We show five main experimental results of drying: Absorbance spectrum broadening (Figure 2a), luminescence spectrum red-shift (Figure 2b), decrease in fluorescence (Figures S5b and S5d), life-time shortening (Figure 2c) and changes in life-time component weights (Figure S6). We also show that luminescence peak order of PC and APC, on an energetic scale, is reversed in the dry phase (Figure 2b). In order to address the experimental data we developed a theoretical framework based on averages from stochastic realizations of chromophore energies, dressed by appropriate homogeneous line-shape functions. This framework is able to explain our results, taking into account chemical and biological constraints.

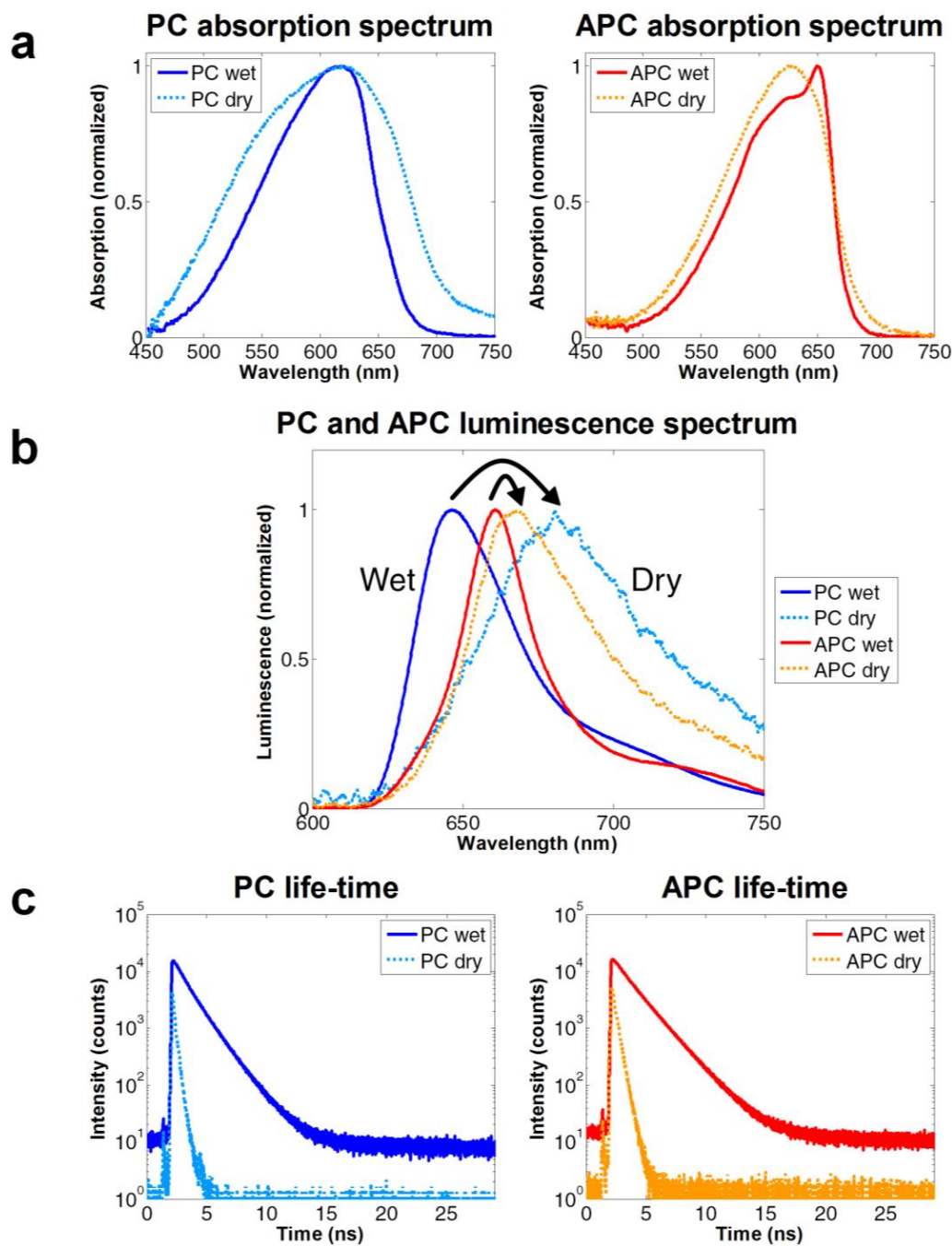

**Figure 2.** Optical measurements. (a) Normalized absorption spectra of PC (left) and APC (right) in wet and dry phases. The dry spectrum is broader than the wet spectrum for both proteins. (b) Normalized luminescence spectra of PC and APC in wet and dry phases. Both proteins are red-shifted in the dry phase. A stronger red-shift is observed in the PC dry phase, reversing the energy flow direction from APC to PC rather than from PC to APC as in the wet phase. (c) Life-time measurements of PC (left) and APC (right) in wet and dry phases. In both cases, life-times are shorter in the dry phase.

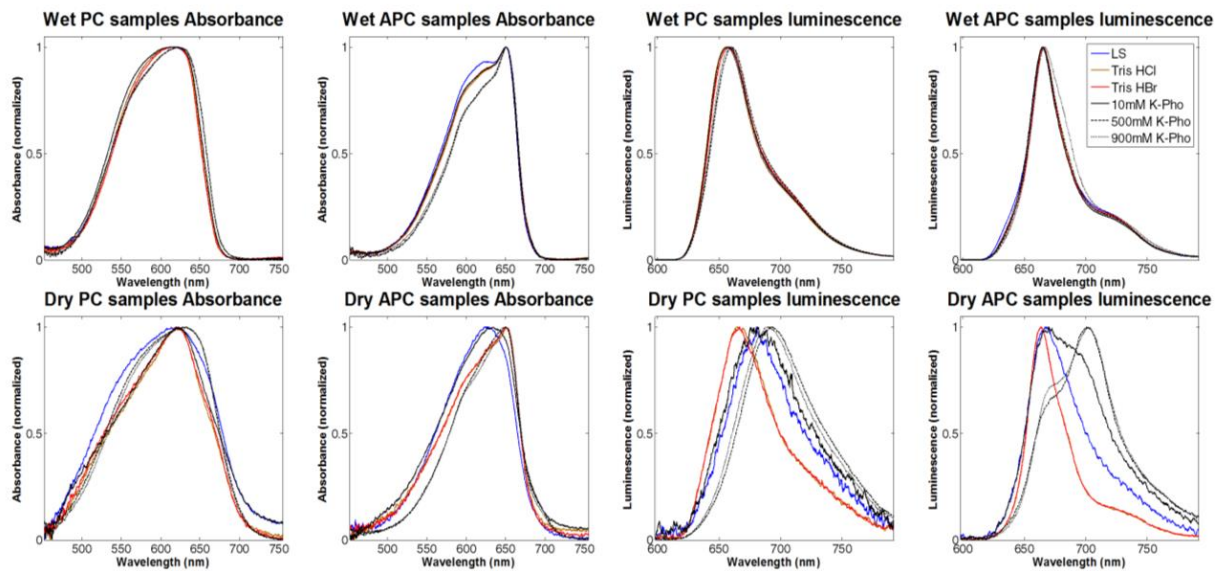

**Figure 3.** Buffer influence. Absorption spectra and luminescence spectra comparison of PC and APC in the wet phase and in the dry phase. PC and APC solutions were prepared in 6 different buffers. Each graph shows the differences between the 6 buffers. Blue curves represent low salts (LS) buffer that was dialyzed against DDW, and its salt concentration is lower than 25 μM. Brown and red curves represent 10 mM Tris HCl and Tris HBr buffers, respectively. Continuous, dashed and dotted black curves represent potassium phosphate buffer at 10 mM, 500mM and 900 mM, respectively. In all cases the variance between different buffers becomes prominent in the dry phase. The wet phase was taken at the beginning of the process.

## THEORY

The theory models the absorption and luminescence spectra, constrained by the weights obtained from life-time traces. We were required to perform averages in order to account for differences within the ensemble measured (inhomogeneous broadening), while each individual chromophore presents a homogeneous line-shape function which takes into account a zero-phonon line and a vibrational progression. The result of this procedure is shown in Figures 4, S2 and S4. Care was taken in finding a set of parameters which would require the least additional tuning for quantities that should not vary significantly from wet to desiccated samples.

Since temperature is a very important quantity for the determination of the homogeneous line-shapes, and given that the temperature does not vary considerably in our experiments, we implemented a model where the homogeneous line-shapes for each chromophore are the same for either wet or desiccated conditions. Moreover, these line-shapes are almost unchanged for every buffer utilized. On the other hand, since the process of desiccation will change the polarity of the medium, it is expected to change the average electric field around the chromophores and thereby produce a change in the individual chromophore's excited state energy. When desiccated, the shift of individual chromophores was found to lie between 300-500 cm$^{-1}$, with a magnitude that is in full agreement with previous chromatic shift studies.[36] These shifts were important to match the desiccated samples absorption spectra (Figure S2) but since they were never greater than the energy difference between chromophores, the hierarchy of energies in ascending order β84, α84 and β155 was maintained.

Shifts of energy alone cannot account for the single peaked shape of absorption from desiccated samples, since the narrower contributions from wet samples shown in Figures 4a and 4b, start to reflect their individual structure when separated. This fact, together with the observation that desiccated samples present a broadening at both lower and higher energy tails with respect to the wet absorption spectra, can be accounted for by increasing the inhomogeneous broadening substantially. The fact that the desiccated spectra exhibits a greater broadening at higher energies as opposed to lower energies, is a result of the inhomogeneous broadening of the β155 chromophore, which is greater than that of the α84 or β84 chromophores

The inhomogeneous contribution, the energies of the chromophores and the homogeneous line-shape used for absorption spectra calculations, were used to calculate the luminescence spectra, shown in Figures 4c, 4d and S4. For this calculation we must account for the equilibration of the excited manifold, which introduces reorganization energy and specific weights of excitonic eigenstate populations, proportional to the Boltzmann factor associated to their energy for the specific realization of inhomogeneous noise. The eigenstates are calculated based on a Hamiltonian that spans the Hilbert space of the hexameric structure for both wet and desiccated samples. Furthermore, in order to obtain agreement with the experiment, and to help address the quenching chromophores, each chromophore contribution to the luminescence spectra was weighted by a factor Φ. The factor Φ, the average populations from the diagonalization of the hexamer Hamiltonian weighted by Boltzmann factors, and, the same homogeneous line-shape functions used for absorption spectra calculations, were used to simultaneously match luminescence spectra and the amplitudes of the contributions found for the life-time traces. This procedure accounted for the vanishing luminescence from the β155 chromophores, a very minor luminescence from α84 (~7%), and a major luminescence from β84 chromophores (~93%). The same procedure was performed to obtain APC absorption and luminescence spectra, with the results presented in Figures S2c, S2d, S4c and S4d.

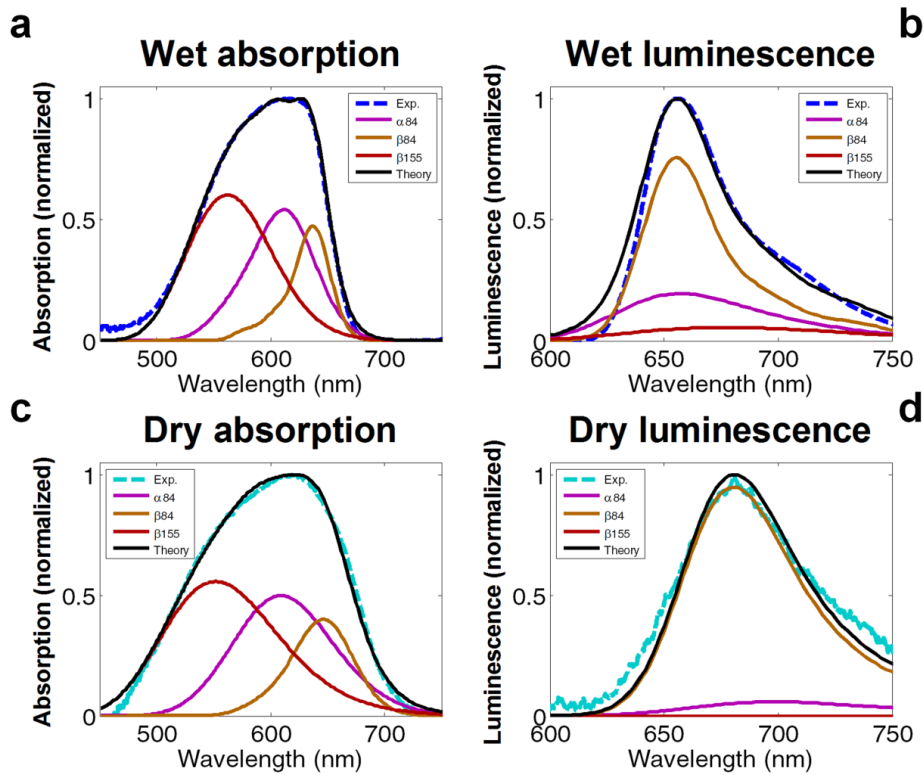

**Figure 4.** Results of the theoretical model for the description of experimental spectra of PC. Figures (a) and (b) present the wet phase model absorption and luminescence spectra, respectively. Figures (c) and (d) present the dry phase LS model for absorption and luminescence spectra, respectively. The experimental result is shown in colored dashed lines, while the results of the α84, β84 and β155 contributions are presented in continuous purple, brown and red lines, respectively. The black lines are the sum of the three contributions.

## DISCUSSION

We suggest that the spectra associated with each phase indicate a specific functionality of the antenna changing as a result of reordering and environmental conditions. In the wet phase, chromophore energy levels are known to be in descending order, β155 → α84 → β84 for PC and α84 → β84 for APC.[26,37–39] Therefore excitons will be transferred into the inner chromophores[40] (Figure 5a). As a consequence most of the luminescence is emitted from the inner chromophore, β84.

However, in the dry phase the broader absorption spectra are compatible with a stronger exciton-environment and adjacent protein interactions. This increases the inhomogeneous broadening in the transition energies of the chromophores[41] (Figure 5b). Since emission occurs after thermal equilibration,[42] luminescence red-shift is expected.

Heuristically, it is reasonable to expect that desiccation affects the external chromophores more drastically due to their greater exposure to the solvent and the fact that they are influenced by its salt concentration.[26,43] To strengthen this claim we show that the presence of different ions in the solution strongly impacts the energy levels in the dry phase (Figure 3). Thus, we would expect that external chromophores will present greater broadening (Figures S2b and S2d and Table S3). Indeed, APC, which lacks the most external chromophore, β155, shows less broadening than PC. Interactions with neighboring proteins also affect the external chromophore environment, making it more similar to the internal part of the PBP. This modified band structure enables the reversing of energy flow direction.

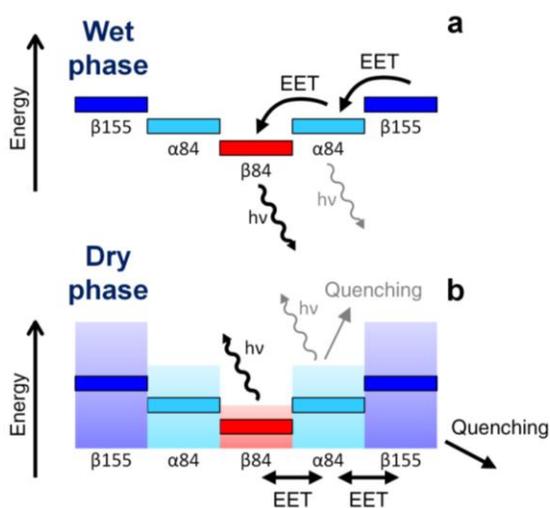

**Figure 5.** PC energy diagram. (a) Wet phase model - PC chromophores funnel energy towards PC center (β84) thanks to gradual order. Luminescence is emitted from both β84 and α84, however the luminescence from the former is stronger. (b) Dry phase model - when PC is dried, energy bands are broadened, especially the two externals. Excitons can migrate in both directions due to strong coupling. This disrupts the directionality characteristic of the wet phase and enables energy quenching to the environment mainly from β155, as well as from α84. Luminescence still emanates from the two inner chromophores. In this phase most of the energy is dissipated outwards to the environment, hence reversing the energy flow. In contrast to the wet phase, where PC hexamers have greater spatial separation, in the dry phase the proteins aggregate and the external chromophores interact strongly with vibronic states, allowing non-radiative decay of energy to the environment as seen by the decrease in luminescence (Figures S5b and S5d). This energy 'leakage' to the environment is more prominent in the outer chromophores, and together with their increased inhomogeneous broadening, explains the enhanced coupling to the environment.

Figure 5 summarizes the main points of the model. The different degrees of inhomogeneous broadening, being large for the externally exposed chromophores and progressively smaller for the internal ones, randomize EET directions. Good coupling between external chromophores and the environment supports efficient quenching. All buffers studied present spectra that are consistent with this energy flow inversion, as quantified in Table S4. Life-times and components weights are discussed in the Supplementary Information section. It is interesting to note that quenching in PBS in the absence of OCP has also been suggested recently by Gwizdala *et al.*.[44]

Furthermore, this model can also provide insights into PBS dynamics *in-vivo* and their effect on the function of the cyanobacterial photosynthetic apparatus while undergoing reorganizational changes such as in desert sand crust cyanobacteria.[35] When efficient transfer is required, energy is transferred from PC's external chromophores to its inner chromophores and proceeds through the center of the PC's rod towards the APC core (Figure 6a). The fact that APC's luminescence peaks at lower energy than PC's creates an EET directionality from rods to core. The inner part of the rod is less exposed to the environment and therefore is preferred as an isolated channel for EET.

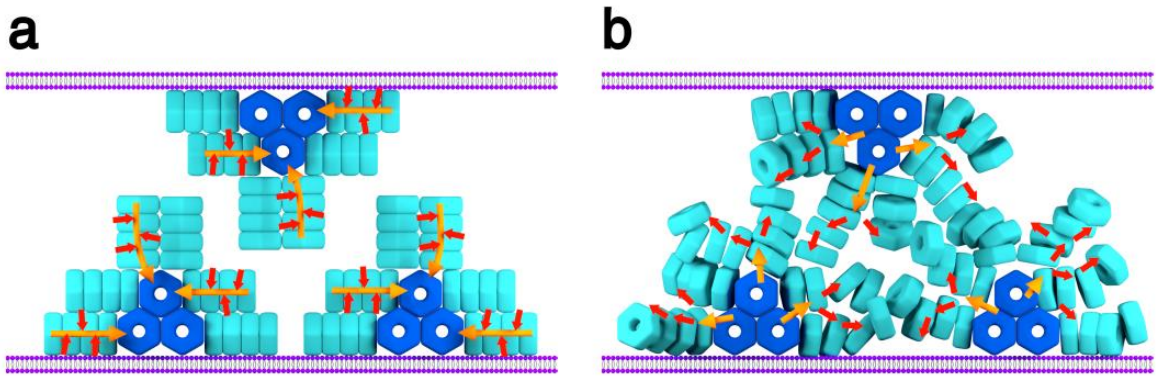

**Figure 6.** A Model for the possible effects of aggregation state on energy transfer paths in PBS. (a) In the organized phase, energy is funneled towards RCs. Photons are absorbed at PC hexamers and create excitons. Excitons are transferred from external chromophores to inner chromophores, propagate along the rod to the APC core and finally reach the RC. (b) In an aggregated phase, PC hexamers are detached from the core and organized randomly. Energy direction flow is reversed and flows from APC core to PCs.

However, when energy quenching is required we suggest that the PC rod hexamers lose their organized structure (Figure 6b). This aggregation of PCs may be similar to the *in-vitro* dry phase. In this phase the aforementioned energy band broadening occurs and allows for the required energy quenching. The reorganization of PCs reverses the energy flow in two ways: Energy is transferred from APCs to PCs (Figure 2b); and from the inner chromophores of PC to its external chromophores (Figure 5b). Transfer of energy to the external band is essential for heat dissipation due to the strong coupling to the environment or to electronic trapped states. Our model seems to explain all experimental observations. Nevertheless other models should be suggested and considered. Energy quenching without the presence of carotenoid may stem from different mechanisms such as conformational changes and ions influence[45–47]. These explanations are not based on the coupling to the environment or to neighboring proteins. Future experiments may teach us whether the quenching mechanism is based on solution concentration solely, or utilizes the coupling to external systems as well.

Lastly, it is interesting to note that PBSs are unique among antenna complexes in the large distances between chromophores generating an intermediate coupling regime. Therefore, small changes in conformation can generate large effects on the energetics of the EET process, as our model suggests. The exposure of chromophores to the environment and their interactions with adjacent proteins are shown to tune the cyanobacterial light harvesting system between two modes: an efficient funnel towards RCs or an efficient quenching to the environment away from RCs. This reversible mechanism may be applicable to cyanobacterial light harvesting systems and possibly to other photosynthetic organisms as well that are subject to significant fluctuating environmental conditions.

## AUTHOR INFORMATION

Corresponding Author
Prof. Yossi Paltiel
Email: paltiel@mail.huji.ac.il
Tel: +972-2-6585760
Author Contributions
D.H. isolated and produced the proteins. I.E. performed the optical measurements and the data analysis. F.C.S. developed the theoretical model. Y.P., N.K., N.A., S.Y. and M.B.P. supervised this research. All authors were involved in writing and proofing the manuscript.

## SUPPORTING INFORMATION

Additional results, figures, tables and theory are included in the Supporting Information.

## ACKNOWLEDGMENT

This research was supported by a grant no. 3-12405(151235) from the Ministry of Science, Technology and Space, Israel, the National Science Council (NSC) of Taiwan, the Israel Science Foundation (1576/12 and 843/16), the US-Israel Bi-National Science Fund (2014395), the ERC Synergy grant BioQ, the EU STREP project PAPETS and an Alexander von Humboldt Professorship. We also want to thank Ms. Lior Bezen for her assistance in life-time measurements.

## ABBREVIATIONS

PBS, phycobilisome; PC, phycocyanin; APC, allophycocyanin; PBP, phycobiliprotein; PCB, phycocyanobilin; OCP, orange carotenoid protein; RC, reaction center; EET, excitation energy transfer; LS, low salts; TCSPC, time-correlated single-photon counting.